\documentclass[prd,
twocolumn,showpacs,preprintnumbers,amsmath,amssymb,
a4paper,nofootinbib,longbibliography]{revtex4-2}

\usepackage{setspace}

\usepackage{nicefrac}

\usepackage{float}

\usepackage{esint}

\usepackage{empheq}
\usepackage[most]{tcolorbox}

\usepackage{natbib}
\usepackage{amssymb,amsbsy,amsmath,amsfonts}
\usepackage{graphicx}
\usepackage{epsf,epsfig,float,latexsym,amsthm,fancyhdr,rotating}
\usepackage{graphics,psfrag,longtable}
\usepackage{slashed}
\usepackage[normalem]{ulem}
\usepackage[colorlinks,citecolor=blue,linktoc=all,linkcolor=cyan,urlcolor=blue]{hyperref}
\usepackage{upgreek}

 \usepackage{multirow}

\def\beq{\begin{equation}}
\def\eeq{\end{equation}}
\def\bea{\begin{eqnarray}}
\def\eea{\end{eqnarray}}
\def\beqa{\begin{equation}\begin{array}{l}}
\def\eeqa{\end{array}\end{equation}}
\def\eqlab#1{\label{eq:#1}}

\def\seclab#1{\label{sec:#1}}
\def\eref#1{(\ref{eq:#1})}
\def\Eqref#1{Eq.~(\ref{eq:#1})}

\def\secref#1{Section \ref{sec:#1}}

\def\half{\mbox{$\frac{1}{2}$}}

\def\barr{\left(\begin{array}{c}}
\def\earr{\end{array}\right)}
\def\bmat{\left(\begin{array}{cc}}
\def\emat{\end{array}\right)}
\def\al{\alpha}

\def\ga{\gamma} 
 
  \def\eps{\epsilon}

\def\si{\sigma}

\def\nn{\nonumber}
\def\dd{\mathrm{d}}

\DeclareMathOperator\im{Im}

\def\3d{3-D}

\def\piEFT/{$\slashed{\pi}$EFT}

\def\2PE{2$\upgamma$}

\usepackage{xcolor}
\usepackage{bm}
\usepackage{slashed}
\usepackage{graphicx}
\allowdisplaybreaks

\makeatletter
\g@addto@macro\bfseries{\boldmath}
\makeatother

\begin{document}

\author{Vladimir Pascalutsa}
\affiliation{Institut f\"ur Kernphysik,
 Johannes Gutenberg-Universit\"at  Mainz,  D-55128 Mainz, Germany}

\title{New superconvergence relations for spin and tensor structure functions  of $\gamma\gamma$ fusion}

\begin{abstract}
The Burkhardt--Cottingham sum rule is an exact superconvergence relation for a spin-structure function, derived from general principles of light absorption and scattering, and valid at any momentum transfer $Q^2$. I illustrate
how a class of such relations emerges from the Siegert point, an unphysical kinematical point where both the probe and the target are at rest.
From light-by-light scattering,  new sum rules for $\gamma^\ast \gamma^\ast$ fusion are emerging, valid for arbitrary photon virtualities.  Regarding the convergence of these relations, there is a simple argument for the suppression of longitudinal photon polarizations at high energy. Among its consequences is the prediction of $\sigma_L/ \sigma_T  \to 0$ at high energy, for the ratio of  unpolarized nucleon photoabsorption cross sections. 
\end{abstract}

\date{\today}

\maketitle

\section{Introduction}

The celebrated Burkhardt-Cottingham (BC) sum rule \cite{Burkhardt:1970ti} is
a superconvergence relation,
\beq
\eqlab{BCg2}
\int_0^1 \dd x\, g_2(x,Q^2) = 0,
\eeq 
for one of the two spin-structure functions $g_{1,2}(x,Q^2)$  of a target particle 
probed by a virtual 
photon with energy\footnote{More specifically, the photon with 4-momentum $q$ is absorbed by a target particle with 4-momentum $p$, with kinematic invariants given by: $\nu= p\cdot q/M$, $p^2=M^2$, $q^2 = -Q^2$, using the ``mostly-negative" Minkowski metric and natural units.} $\nu = Q^2/(2Mx)$ and virtuality $Q^2$; $M$ is
the particle mass, $x$ is the Bjorken variable. 
Derived from the general principles of causality (analyticity), unitarity (optical theorem), 
Lorentz symmetry, and the electromagnetic gauge invariance, manifested in the process of photon absorption and scattering (see, e.g., \cite{Lampe:1998eu,Drechsel:2002ar, Pascalutsa:2024hvy}), it provides an exact constraint, valid at any $Q^2$ (up to possible convergence issues, cf., e.g., \cite{Jaffe:1989xx}).
 
 The BC sum rule has been tested experimentally at various $Q^2$ on proton, neutron, 
deuteron, and $^3$He targets and found to hold within uncertainties~\cite{Amarian:2002,Prok:2014,Slifer:2010}.
It is applicable to any particle with spin, albeit mainly
used in the context of nucleon and nuclear structure \cite{Kuhn:2008sy,Deur:2018roz}, and is particularly useful in the evaluations of the nuclear effects in the hyperfine splitting of hydrogen-like atoms, cf.\ \cite{CLAS:2021apd,Antognini:2022xoo,Ruth:2024bsl} for recent developments.

The present paper shows that a broad class of such relations
can be obtained from dispersion relations for transverse ($T$) and longitudinal ($L$) Compton amplitudes, by using the equivalence of these amplitudes ($T=L$) at the Siegert point~\cite{Siegert:1937yt}:
$\nu^2 = -Q^2$. This equivalence is a manifestation of the fact that, at this unphysical kinematical point, both the target and the photon are at rest (photon momentum $|\vec q\,|  = \sqrt{Q^2+\nu^2} = 0$), and hence, longitudinal and transverse polarizations are indistinguishable. I shall refer to it as the ``LT-degeneracy" at the Siegert point, for lack of a better name in the literature.\footnote{In Ref.~\cite{Biloshytskyi:2023fyv}, we called it the \emph{Siegert theorem} — but, strictly speaking, it isn’t. In retrospect, it may be viewed as a lemma to the actual theorem~\cite{Siegert:1937yt} which, in addition, exploits the current conservation,  multipole expansion, the extreme non-relativistic (static) approximation, etc.; see e.g.,~\cite{Friar:1984zza} for details.} 
The LT-degeneracy is 
 correlated with the spin of the target particle, because 
 the total helicity is conserved.

This paper is organized as follows. 
\secref{HEP} presents a simple argument of why the Compton amplitudes and photoabsorption cross sections for longitudinal photons 
are suppressed at high energy and discusses its most evident implications. In \secref{unpol}, we begin to 
consider the forward Compton scattering at the Siegert
point, focusing at first on the unpolarized case.
\secref{BCsec} we consider the spin-dependent amplitude and
rederive the BC sum rule, emphasizing the difference with the standard derivation. \secref{LbL} is devoted to the light-by-light scattering and presents the new relations for the structure functions of
$\ga\ga$ fusion. \secref{conclusion} contains conclusions and an outlook
to near-future applications.

\section{High-energy suppression of longitudinal versus transverse components}
\seclab{HEP}
On the question of UV-convergence, it is important to realize that 
the longitudinal amplitudes and cross sections must be suppressed with energy, as compared to their transverse counterparts. 
The suppression factor is roughly $1/\nu$, per longitudinal photon.  
This is because in the limit $\nu \to \infty $, just as 
for  $Q^2 \to 0 $, the photon becomes real.
Indeed, the virtual-photon velocity is the ratio of photon momentum to its energy (in natural units):
\beq
v_{\gamma^\ast} = \frac{\sqrt{Q^2+\nu^2}}{\nu} \stackrel{\nu\to \infty}{\to } 1.
\eeq 
and the longitudinal components will ultimately  decouple.
This is also true in the Bjorken limit: fixed $x$, large $\nu$ and $Q^2$.

This suppression is automatically ensured by the electromagnetic gauge invariance.\footnote{Consider a typical matrix element, $\epsilon_L \cdot J$, where $$
\epsilon_L = 1/Q (|\vec q\,|, 0 , 0 , \nu) $$
is the photon polarization 4-vector in the lab frame, and $J$ is a conserved current. Note that it is singular
in the limit of $Q\to 0$ or $\nu\to \infty$. However, thanks to gauge-invariance, 
$$q\cdot J =0 = \nu J^0 - |\vec q\,| J^3, $$
and the factors are turning up-side-down:
$$
\epsilon_L \cdot J = 1/Q\,\big(|\vec q\,|  J^0 -  \nu J^3\big) = (Q/\nu) J^3. $$
This agrees with the decoupling of longitudinal components, and also explicitly shows why they are accompanied by  a factor of $Q/\nu$.}
Yet, it is often missed in assessments of the Regge behavior
at large $\nu$. 
 In particular, 
the asymptotic Regge  behavior of the unpolarized cross sections 
of total-photoabsorption on the proton, the transverse $\sigma_T$ and longitudinal $\sigma_L$, 
is often assumed to be the same in DIS fits aimed at 
the extraction of unpolarized structure functions ($F_1 \sim \sigma_L$ and $F_2 \sim \sigma_T+\sigma_L$), see, e.g., \cite{Alwall:2004wk,Gao:2017yyd} and references therein.
In reality, where the electromagnetic gauge invariance is exact, $\sigma_L$ should be suppressed by extra powers of $1/\nu$, to ensure
that it vanishes in the real-photon limit of $\nu\to \infty$.

This becomes apparent in any simple model with gauge invariance. For instance, in the naive parton model, one has \cite{Feynman:2019rot}: 
\beq
\sigma_L = \frac{Q^2}{\nu^2}\sigma_T\, .
\eeq 
We have verified the suppression in other simple examples~\cite{Biloshytskyi:2024nkj}; first of all, in 
the leading-order QED, where the photoabsorption cross-sections
are given by tree-level Compton scattering. In this case, the polarized cross sections happen to exactly follow this trend.\footnote{Note that the helicity-difference cross-section $\sigma_{TT} =\half (\sigma_{1/2} - \sigma_{3/2})$ is not positive definite.  The full QED expressions can be found in Ref.~\cite{Hagelstein:2017obr} and references therein.} 
\beq
\eqlab{helimit}
\si_{LT}(\nu, Q^2)  \stackrel{\nu \to\infty}{\to} -\frac{Q}{\nu} \si_{TT} (\nu, Q^2).
\eeq 
This feature is, of course,  crucial in assessing the convergence of 
the relations discussed below. 

The specific power of suppression may vary from theory to theory, but it is clear that
\beq
\si_L(\nu, Q^2) \stackrel{\nu \to\infty}{\to} 0,
\eeq
and this is true for any cross section of longitudinal-photon absorption.
For the proton, this means, given the rising $\sigma_T^p$~\cite{PDG:2024}, that the ratio of longitudinal
and transverse cross sections must vanish:
$\si_L^p/\sigma_T^p \to 0$, for $\nu \to\infty$.

\section{Virtual Compton scattering at the Siegert point}
\seclab{unpol}
The forward doubly-virtual Compton scattering (VVCS) on a spin-1/2 target, such as the nucleon, can be described by 
four scalar amplitudes: two unpolarized (or, spin-independent):
$T_1(\nu,Q^2)$ and $T_2(\nu,Q^2)$
and two polarized (or, spin-dependent):
$S_1(\nu,Q^2)$ and $S_2(\nu,Q^2)$, see 
\cite{Drechsel:2002ar,Hagelstein:2015egb} for further details
and notations. In this section, we consider the unpolarized case.

The optical theorem relates the discontinuity of these amplitudes to the structure functions, e.g.,  $\im T_{1,2}(\nu, Q^2) \sim F_{1,2} (x, Q^2) $. The amplitude $T_1$ is purely transverse, whereas the purely longitudinal one is given by the following linear combination
of $T_1 $ and $T_2$:
\beq
T_L (\nu, Q^2)  = \Big( - T_1 + \frac{\nu^{2}+Q^2}{Q^2} T_2
\Big) (\nu, Q^2).
\eeq 
 From this expression it is immediately clear that, at the Siegert point ($|\vec q | = 0$), $T_L=-T_1$, and hence the `LT degeneracy'. Assuming
\emph{unsubtracted} dispersion relations for both $T$ and $L$ 
amplitudes, with the corresponding optical theorem,
\begin{subequations}
\bea
\im T_1 (\nu, Q^2)&=& \nu\,  \sigma_T(\nu, Q^2),\\
\im T_L (\nu, Q^2)&=& \nu \, \sigma_L(\nu, Q^2),
\eea
\end{subequations}
we would obtain:
\beq
\eqlab{BCunpol}
0 = \frac{2}{\pi}\int_{\nu_\mathrm{el}}^\infty \dd \nu\frac{\nu^{2}}{\nu^{2}+Q^2}\left[\sigma_T(\nu,Q^2) + \sigma_{L}(\nu,Q^2)\right],
\eeq 
where $\nu_\mathrm{el} = Q^2/2M$ is the elastic threshold.
Of course, this cannot hold, since both cross sections are positive-definite. There is also a convergence issue. According to our gauge-invariance argument, the $T$- and $L$-cross sections do not scale with the same power of $\nu$: $\sigma_L$ is suppressed. The $T$-term alone is therefore unlikely to converge, as is also known empirically for the nucleon from the Regge behavior of hadronic cross sections~\cite{Donnachie:1992}.

Both issues are resolved by using a subtracted dispersion relation for the transverse amplitude,
\beq
T_{1}(\nu,Q^2) = T_{1}(0,Q^2) + \frac{2\nu^2}{\pi} \int_{\nu_\mathrm{el}}^\infty \dd \nu’ , \frac{\sigma_{T}(\nu’,Q^2)}{\nu^{\prime,2}-\nu^2},,
\eeq
where $T_{1}(0,Q^2)$ is the subtraction function. This function is often viewed as a nuisance, most notably in calculations of nuclear-polarizability effects in muonic atoms (see, e.g.,~\cite{Pohl:2013yb,Carlson:2015jba,Antognini:2022xoo} for reviews).

The BC sum-rule analogue, written in \Eqref{BCunpol}, then becomes
the sum rule for the subtraction function \cite{Gasser:2015dwa,Hagelstein:2020awq,Biloshytskyi:2023fyv}:
\bea
T_1(0, Q^2) &=& \frac{2}{\pi}\int_{\nu_\mathrm{el}}^\infty \dd \nu\frac{\nu^{2}}{\nu^{2}+Q^2}\left(\frac{Q^2}{\nu^2} \sigma_T - \sigma_{L}\right)(\nu,Q^2),\nn\\
 &=& -\frac{4\pi \al}{M}\int_0^1 \frac{\dd x}{x^2} \,F_L(x,Q^2),
\eea
where in the second line, $F_L=F_2-2xF_1$ is the longitudinal structure function, obtained by using
\begin{subequations}
\bea
\frac{4\pi^2 \al}{M} F_1(x, Q^2) &=& \nu\,  \si_T(\nu, Q^2), \\
\frac{4\pi^2 \al}{\nu} F_2(x, Q^2) &=& \frac{\nu\,  Q^2}{\nu^2+Q^2} \si_T(\nu, Q^2) ,
\eea
\end{subequations}
with $\al$ the fine-structure constant.

The integrals look dangerously singular in 
the limit of high energy $\nu$ or low $x$.
Certainly the first term, containing $\sigma_T$, 
is known to be rapidly convergent, as it essentially gives the Baldin sum rule \cite{Baldin:1960}. The second term may also converge if each longitudinal photon indeed carries a suppression factor of $Q/\nu$,
suppressing $\sigma_L$ by an extra $Q^2/\nu^2$. 
In simple examples the convergence is seen explicitly, see Ref.~\cite{Biloshytskyi:2023fyv}.

To summarize, in the unpolarized case, one does not obtain a superconvergence relation
at the Siegert point, but rather a sum rule for the subtraction function $T_1(0,Q^2)$. If convergent for the nucleon, the resulting relation may still be useful for
a data-driven determination of the nucleon polarizabilities and their effect
in the Lamb shift of hydrogen-like atoms.

\section{Derivation(s) of Burkhardt-Cottingham sum rule } \seclab{BCsec}

Let us now turn the attention to the spin-dependent case, described
by VVCS amplitudes $S_1$ and $S_2$.

The BC sum rule is derived from a dispersion relation 
for $S_2$, in the limit $\nu \to 0$. One first assumes 
an unsubtracted dispersion relation for $\nu S_2(\nu, Q^2)$,
and then requires $S_2$ to have no pole at 0, i.e.:
\beq
\eqlab{limitS2}
\lim_{\nu\rightarrow0} \nu\, S_2(\nu,Q^2)=0.
\eeq
This derivation seems somewhat arbitrary. In particular, it is
not very clear why one starts with the dispersion relation
for $\nu S_2$, and not simply $S_2$. In the latter case, we would not have the BC sum rule. One argument 
is that (the Born contribution to) $S_2$, in the limit of $Q^2\to 0$, does have a pole at $\nu=0$, 
which would invalidate the dispersion relation.
Hence, it makes sense to consider $\nu S_2$. However, if there is a pole, why can we demand \eref{limitS2}?
Rather than struggling with these questions, we take an alternative, more straightforward path.

Consider the spin-dependent amplitudes corresponding to $T$ or $L$ photon polarizations, expressed as a linear combination of original amplitudes as:
\begin{subequations}
    \bea
S_{TT}(\nu,Q^2)&=&\frac{Q}{M}\!\left[\frac{\nu}{Q}S_1(\nu,Q^2)-\frac{Q}{M}S_2(\nu,Q^2)\right]\!,\qquad\\
S_{LT}(\nu,Q^2)&=&\frac{Q}{M}\!\left[S_1(\nu,Q^2)+\frac{\nu}{M}S_2(\nu,Q^2)\right]\!.
    \eea
\end{subequations}
We assume that these amplitudes satisfy the unsubtracted dispersion relations:
\begin{subequations}
\eqlab{DRs}
    \bea
S_{TT}(\nu,Q^2)&=&\frac{2\nu}{\pi}\int_{\nu_\mathrm{el}}^\infty \dd \nu' \frac{\nu' \sigma_{TT}(\nu',Q^2)}{\nu^{\prime\,2}-\nu^2},\\
S_{LT}(\nu,Q^2)&=&\frac{2}{\pi}\int_{\nu_\mathrm{el}}^\infty \dd \nu' \frac{\nu^{\prime\,2} \sigma_{LT}(\nu',Q^2)}{\nu^{\prime\,2}-\nu^2},
    \eea
\end{subequations}
where  we have used the optical theorem,\footnote{Sometimes the amplitudes $S_{TT}$ and $S_{LT}$ are denoted as $g_{TT}$ and $g_{LT}$, respectively (e.g., Ref.~\cite{Drechsel:2002ar}).
Also, because of the ambiguity of defining  flux factor of the virtual photon, a different phase-space factor (which multiplies the cross section) can be used in the optical theorem; e.g., many  use $|\vec q\,|$ instead of $\nu$ (e.g., \cite{Biloshytskyi:2023fyv}). However, the expressions in terms of structure functions are independent of these conventions.} $\im S_i (\nu,Q^2)=\nu \, \sigma_i(\nu,Q^2)$ to relate the amplitudes to the longitudinal-transverse ($i=LT$) and the transverse-transverse ($i=TT$) photoabsorption cross sections. 
Note that convergence of these dispersion relations is warranted by convergence of, respectively, the GDH \cite{Gerasimov:1965et, Drell:1966jv} and Schwinger \cite{Schwinger:1975uq,HarunarRashid:1976qz} sum rules, which schematically read as:
\begin{subequations}
\eqlab{schem}
\bea
S_{TT}(\nu\to 0,0) &:&  \varkappa^2 \sim \int \frac{d\nu}{\nu} \si_{TT}(\nu, 0) \\
\frac{S_{LT}(0,Q^2)}{Q}\Big|_{Q\to 0 } &:&  \varkappa \sim \int \dd\nu \frac{\si_{LT}(\nu, Q^2)}{Q}\Big|_{Q\to 0 },
\eea 
\end{subequations}
where $\varkappa$ is the anomalous magnetic moment. Here 
it is important to emphasize that, as argued above,  a longitudinal 
photon brings in an extra $1/\nu$ at high energy, see \Eqref{helimit}.
Hence the convergence of the two dispersion relations is correlated.

The second step of our derivation is the Siegert point ($\nu = i Q$), where we have an equivalence of the $T$- and $L$-amplitude:
\beq
\eqlab{Siegert}
-i S_{TT}(iQ,Q^2)=S_{LT}(iQ,Q^2).
\eeq
This is a manifestation of the fact that, at this kinematical point, the photon is at rest, its momentum $\vec q  = 0$ , and hence longitudinal and transverse polarizations are indistinguishable.

Plugging in here the dispersion relations of \Eqref{DRs}, we find
the BC sum rule:
\beq
\eqlab{BCalt}
0=\frac{2Q}{\pi}\int_{\nu_\mathrm{el}}^\infty \dd \nu \frac{\nu }{\nu^{2}+Q^2}\left[\frac{\nu}{Q}\sigma_{LT}(\nu,Q^2)-\sigma_{TT}(\nu,Q^2)\right],
\eeq
which can be cast in its standard form \eref{BCg2} by using: (i) the relation between the spin structure function $g_2$ and the photoabsorption cross sections,
\begin{subequations}
\bea
\im S_2(\nu,Q^2)  &=&  \frac{4\pi^2 \alpha M}{\nu^2} \, g_2(x, Q^2)  ,\\
&=& \frac{M^2\nu}{\nu^2+Q^2}\left[\frac{\nu}{Q}\sigma_{LT}  - \sigma_{TT}\right](\nu,Q^2),\qquad
\eea
and (ii) a change of the integration variable from $\nu$ to  $x = Q^2/2M\nu$.
\end{subequations}

\section{ New sum rules for $\gamma^\ast\gamma^\ast$ fusion }
\seclab{LbL}

A photon by itself can be viewed as the target, probed by a virtual photon via light-by-light (LbL) scattering. In this case, the Siegert point is subtle: one may say it does not exist, since both photons cannot be brought to rest if one of them is real. However, one can start from doubly-virtual forward LbL scattering, characterized by two virtualities, say $q_1^2 = -Q_2^2$ and $q_2^2 = -Q_1^2$, and a crossing-symmetric invariant $\nu = q_1 \cdot q_2$. Note that this $\nu$ differs from the one used above, also dimensionally.

The Siegert point is then defined as:\footnote{Initial steps to study the LbL scattering at this point were undertaken in a PhD thesis~\cite[Sec. 5.4]{Biloshytskyi:2024nkj}.}
\beq
\nu^2_S = Q_1^2 Q_2^2
\eeq
It is interesting that this kinematical point is now real, albeit still unphysical (the total energy, $\sqrt{s} = iQ_1\pm i Q_2$).
This means it lies outside of the physical interval, $\nu_S \notin (\nu_\mathrm{el}, \infty)$, and hence the integrals seen below are not singular.

At the Siegert point the photon four-momenta are orthogonal to any of their polarization vectors:
\beq 
q_i \cdot \eps_j = 0, \quad \mbox{$\forall\,  i,j = 1,2$.}
\eeq
Hence, the Lorenz structure of the doubly-virtual LbL amplitude has only 3 terms, just as for  the real LbL~\cite{Pascalutsa:2010sj}:
\bea
\eqlab{genformLbL}
\mathcal{M}^{\mu_1\mu_2\mu_3\mu_4}(q_1,q_2) & =&  A(\nu_S) \, g^{\mu_4\mu_2} g^{\mu_3\mu_1} \nn\\
&+& B(\nu_S)\, g^{\mu_4\mu_1} g^{\mu_3\mu_2} \\
&+& 
 C(\nu_S) \, g^{\mu_4\mu_3} g^{\mu_2\mu_1}, \nn
 \eea
 with $g^{\mu\nu}$ the metric tensor. 
The LT-degeneracy is hereby immediately exposed: out of the 8 amplitudes needed
to describe the forward-doubly-virtual LbL scattering, only 3  
remain at the Siegert point, as in the case of real LbL.

Therefore, we have 5 inter-relations among the amplitudes, which one can easily derive by contracting \Eqref{genformLbL} with various
polarizations. One of the relations involves the spin structure functions.  It leads to the following superconvergence relation
(in the notation of Budnev \emph{et al.}~\cite{Budnev:1974de}):
\bea
\eqlab{newSR1}
0 &=& \int\limits_{\nu_\mathrm{el}}^\infty \dd \nu \frac{\left[ \tau_{TT}^a -\mbox{$\frac{\nu}{Q_1 Q_2}$} \tau_{TL}^a \right](\nu,Q_1^2,Q_2^2)}{\sqrt{\nu^2-Q_1^2 Q_2^2}} \, ,
\eea 
where $\nu_\mathrm{el} = (Q_1^2+Q_2^2)/2$, and the $\gamma^\ast\gamma^\ast$-fusion 
response functions $\tau_{TT}^a$
and $\tau_{LT}^a$ are analogous to, respectively: $\si_{TT}$ and $\si_{TL}$
response functions considered in VVCS.

We can take the limit of one of the virtualities going to 0, i.e., one of the photons is real. In this limit, the Siegert point is at  $\nu = 0$, i.e.,  the same limit in which the LbL sum rules are normally derived via the low-energy expansion (see, e.g., \cite{Pascalutsa:2010sj,Pascalutsa:2012pr}. 
In this case, the integral of each of the two terms in \Eqref{newSR1} vanishes separately, leading to the
known superconvergence for both spin-structure functions of the (real) photon,
\begin{subequations}
\bea 
\eqlab{BCLbL}
\int_0^1 \dd x\, g_1^\gamma(x,Q^2) &=& 0,\\
\int_0^1 \dd x\, g_2^\gamma(x,Q^2) &=& 0.
\eea 
\end{subequations}
Thus, besides the BC sum rule (second line), we have reproduced the sum rule 
for $g_1^\gamma$, cf.~\cite{Bass:1998bw} and references therein. 

Considering the tensor structure functions, we obtain
another superconvergence relation:
\bea
\eqlab{newSR2}
0 &=& \int\limits_{\nu_\mathrm{el}}^\infty \dd \nu \, \nu \,\frac{\left[ \tau_{TT} -\mbox{$\frac{2Q_1 Q_2}{\nu}$} \tau_{TL} \right](\nu,Q_1^2,Q_2^2)}{\sqrt{\nu^2-Q_1^2 Q_2^2}} \,.
\eea
Just as the generalization of the BC sum rule in \Eqref{newSR1}, 
this relation is valid
for any spacelike virtualities $Q_1^2$ and $Q_2^2$. It will be interesting to extend it to a charged target (with a quadrupole moment) such as the deuteron, as it hints to a sum rule for the quadrupole moment.

\section{Conclusions and outlook}
\seclab{conclusion}
This paper revisits the exact sum rules that follow from general principles of photon scattering and absorption on a target particle with spin. 
Two basic principles are additionally emphasized:
\begin{itemize}
    \item the ``LT-degeneracy'' at the Siegert point, i.e., the equivalence
of the transverse and longitudinal photon polarizations at an unphysical kinematical point,
\item the decoupling of longitudinal photon polarizations in the high-energy limit, $\nu\to\infty$, which should be ensured by the
electromagnetic gauge invariance and unitarity bounds.
\end{itemize}
The Burkhardt-Cottingham sum rule naturally follows from the LT-degeneracy and a few other known and new relations emerge in the same approach.  
In the unpolarized case, it lead to a well-known sum rule for the subtraction function $T_1(0, Q^2)$. 

The case of light-by-light scattering is particularly interesting, as it leads to novel superconvergence relations for $\gamma^\ast\gamma^\ast$-fusion
at arbitrary virtualities, Eqs.~\eref{newSR1} and \eref{newSR2},
which represent the central result of this paper.

These are first relations of this sort, and can, for instance, be used to constrain the doubly-virtual meson transition form factors, similar to the single-virtual case \cite{Dai:2017cvz,Danilkin:2016hnh}.  
In the limit when one of the photons is real, the relation \eref{newSR1} yields the two superconvergence relations for the photon spin-structure functions,
$g_1^\gamma$ and $g_2^\gamma$. 
 
It would be natural to extend this approach to other spin targets. For spin-1 particles such as the deuteron, four tensor structure functions $b_1,\dots,b_4(x,Q^2)$ appear. At present, only the Close–Kumano sum rule is known,
\beq
\int_0^1 \dd x\, b_1(x) = 0,
\eeq
which was shown to hold in the naive parton model~\cite{Close:1987zz}. Whether this relation can be extended to arbitrary $Q^2$ in a model-independent way remains an open question.

The method can also be generalized to higher spins. For spin-3/2, this is of particular interest in the context of the many studies of the $\Delta(1232)$ isobar and other decuplet baryons (see~\cite{Pascalutsa:2006up,Geng:2013xn,Ramalho:2023hqd} for reviews).
Moreover, analogous relations are expected for the gravitational structure functions, which describe a target probed by a graviton rather than a photon~\cite{Polyakov2018,Cotogno2022}.

\acknowledgments
I am grateful to Franziska Hagelstein and Marc Vanderhaeghen for many insightful discussions, checks of the formulae, and a careful reading of the manuscript. I also thank Volodymir Biloshytskyi, Kees Mommers, Sotiris Pitelis, and other participants of the AG Vanderhaeghen seminar for their constructive critique and helpful suggestions.

This work is supported by the Deutsche Forschungsgemeinschaft (DFG) through the Research Unit FOR 5327 “Photon-photon interactions in the Standard Model and beyond” (grant 458854507) and the Collaborative Research Center 1660 ``Hadrons and Nuclei as Discovery Tools'' (grant 514321794).


\bibliography{main.bib}

\end{document}